# D-HAT: a Diatom-inspired structure for a Helmet concept Against Trauma

L. Musenich, L. Strozzi, M. Avalle, F. Libonati[*]

University of Genoa, Polytechnic School, Department of Mechanical, Energy, Management and Transportation Engineering, Via all'Opera Pia 15/A, 16145, Genova, Italy

*Corresponding author: flavia.libonati@unige.it

## Abstract

The primary objective of helmet design continues to be preventing traumatic brain injuries. Yet, achieving an optimal user experience—encompassing aspects such as fit, thermal comfort, breathability, waterproofing, and helmet reusability—is becoming increasingly significant. Thus, designing helmets with multifunctional performance represents the latest technological frontier for these safety devices. This study draws inspiration from the morphology of a specific species of unicellular algae, the *Coscinodiscus sp*. diatom, to design a biomimetic material capable of replicating their cellular structure and multi-functionality. Unlike the biological counterpart, the synthetic material is specially engineered as the inner liner for multi-impact helmets, suitable for urban sports and micro-mobility applications. The material's architecture is modeled using computer-aided design (CAD) tools, and its ability to absorb mechanical energy is analyzed using a numerical and experimental approach based on finite element modeling and quasi-static compression tests conducted on 3D-printed elastomeric samples. Then, through parametric optimization, its performance is maximized. The outcomes demonstrate that the designed material exhibits energy-absorption characteristics comparable to other cellular materials, such as honeycombs, while offering lightweight, breathability, and protection against atmospheric agents.

## Introduction

Helmets represent the most effective technological solution for mitigating the risk of traumatic brain injuries associated with road accidents or sports activities [1], [2]. Their primary function is to protect the user, during impacts, by converting the kinetic energy from collisions into the deformation energy of the helmet materials. Most helmets, regardless their intended use, have a universal basic structure composed of (i) a thin, rigid outer shell and (ii) a thick, more deformable inner liner (Figure 1 (a)) [3]–[9]. The outer shell distributes the impact forces over a larger surface area, preventing high local pressures and protecting against penetration by external objects. Secondary functions include reducing wear-induced abrasion and improving the helmet's aesthetics. Polymeric materials, such as acrylonitrile butadiene styrene (ABS) or polycarbonate (PC), are popular for the outer shell. In contrast, composite materials such as fiberglass, carbon fiber or Kevlar are used for special applications [4], [10]–[12]. The inner liner absorbs and dissipates the energy transferred from the outer shell, thus minimizing head acceleration. This is conventionally achieved using polymeric structural foams such as Expanded Polystyrene (EPS), Expanded Polypropylene (EPP), and expanded Polyurethane (PU) [6], [13]–[15]. These materials are isotropic, easy to manufacture, and possess an impact response curve with a constant stress plateau region (Figure 1 (b)) [16]. This allows them to absorb energy through irreversible deformations (single-impact helmets) or dissipative viscous phenomena (multi-impact helmets), limiting the force transmitted to the user below the skull fracture threshold while increasing displacement.

Given a certain amount of energy to absorb, classic helmet certification standards define design criteria based on the maximum allowable translational accelerations (peak forces) the head can withstand [2], [14], [17]. Originally, helmets were designed empirically by selecting the liner thickness and its mechanical properties to pass the destructive tests required by these standards. The design methodology has been rationalized thanks to analytical models and numerical simulation techniques, improving helmet quality and performance [6], [7], [18]. Recently, research has shown that considering only the translational dynamics of impacts is insufficient for proper helmet design. Impacts often occur at oblique angles, producing translational and rotational inertial forces on the head. Therefore, helmet effectiveness depends on the liner's ability to absorb energy in compression and shear. Common structural foams are not sufficiently deformable under shear loading. Thus, literature has proposed (i) adding rotational damping systems for multidirectional protection or (ii) replacing the foams with 3D-printed architected materials [4], [19]–[25]. Architected materials, specifically, can be custom-designed by controlling both their chemical composition and their lattice-based structure, providing improved impact-attenuation performance compared to stochastic foams [26]–[28]. Their effectiveness has been demonstrated in commercial helmets like HEXR [29] and KOROYD [30] or research concepts [31]–[33]. Yet their potential has only been partially explored. Prevention of traumatic brain injuries remains the primary goal of helmet design. However, an optimal overall user experience, including fit, thermal comfort, breathability, waterproofing, and helmet reusability, is becoming increasingly important [37]–[40], with these design aspects aligning with sustainable development goals [41].

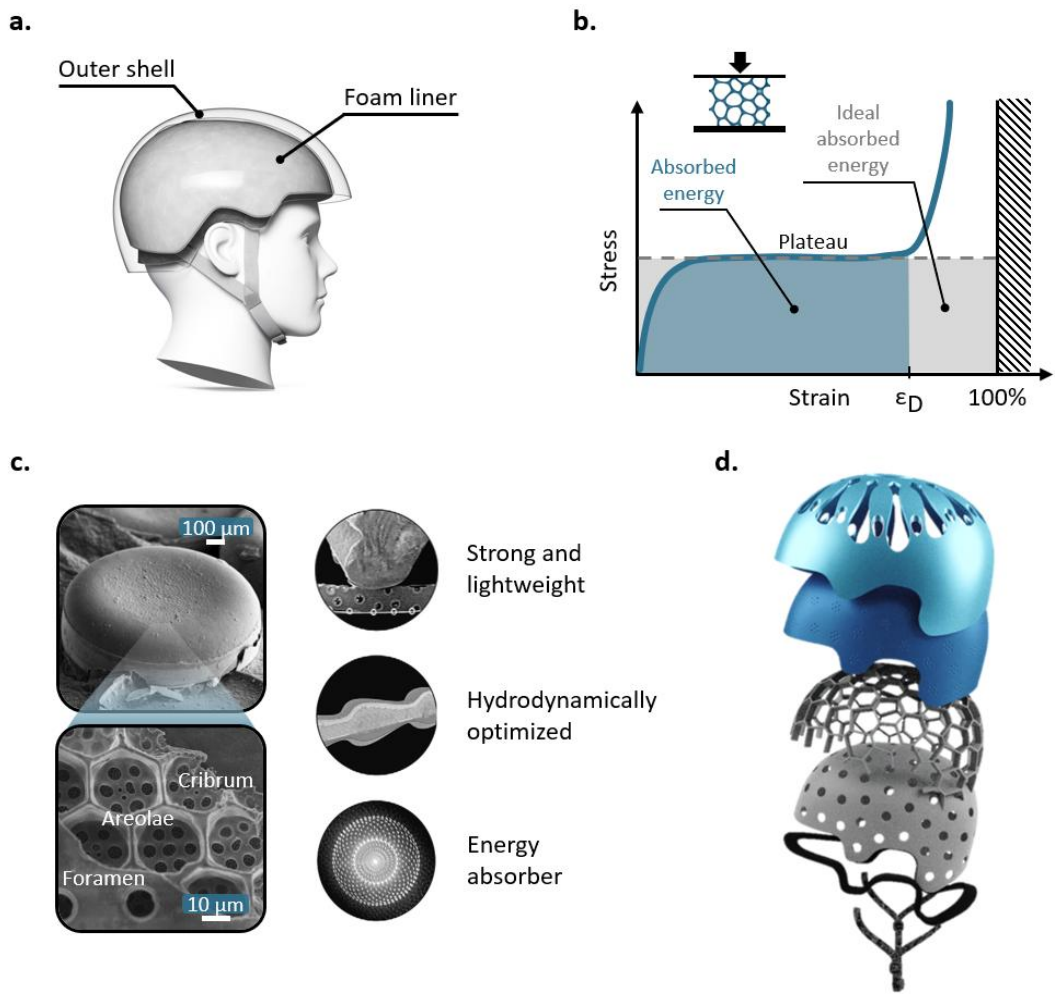

*Figure 1 – (a.) Example of an outdoor activity helmet. The liner must maximize energy absorption per unit volume without exceeding a force that would damage the skull. (b.) Structural foams in helmet liners exhibit a flat plateau-like stress-strain response (solid blue line) during compression, ending at densification ($\varepsilon_D$). The area under the curve represents absorbed energy per unit volume. This behavior tends to the ideal impact response (dashed black line) [16]. (c.) Coscinodiscus sp. diatom frustule at two magnification levels. A close-up shows its multilayer architecture, providing strength, impact resistance, optimal fluid dynamic and optical properties, and lightness (Modified from [34], CC BY 4.0, [35], and [36] with permissions). (d.) The diatom-inspired helmet concept.*

Designing helmets with a multifunctional approach represents the new technological frontier for these safety devices. So how can we create a helmet that is simultaneously safe, custom-made, lightweight, breathable, and with smart capabilities (i.e., self-sensing for structural health monitoring)? Nature can guide us in this endeavor [42]. Biological materials, with their hierarchical organization across multiple length scales, have been optimized over a billion-year long evolution to simultaneously perform multiple functions, combining strength, flexibility, lightness, self-repair, self-sensing, and other smart capabilities [43]–[47]. Transferring these features to engineering materials can be highly rewarding [48]–[52]. An emblematic example of natural multifunctional material is the exoskeleton of diatoms, known as the frustule [36], [53]–[55]. The frustule is an intricate architecture of amorphous biosilica that develops from the nanoscale to the microscale (Figure 1 (c)) [34], [56]–[58]. Despite being composed of an inherently brittle (glassy) material, its hierarchical design confers high mechanical strength and toughness to the overall structure [59]–[63]. This offers the unicellular algae excellent physical protection against predatory attacks or collisions with foreign objects. Additionally, its porous nature allows diatoms to perform photosynthesis while floating in aquatic currents [64]–[67], acquire nutrients through filtration, and expel waste substances without direct contact of the cell with the surrounding environment [68]–[71].

Here, we mimic the frustule of the *Coscinodiscus sp.* diatoms to devise a new liner for an innovative helmet concept (Figure 1 (d)). Starting from the frustule morphology, we create a millimeter-scale biomimetic structure that retains its geometric features but exploits an elastomer as the base material. Its hyperelastic properties modify the natural function of the diatom exoskeleton by giving it multi-impact energy absorption capabilities, without altering the mechanical and fluid dynamic efficiency associated with the frustule topology. Using finite element (FE) analysis, validated by experimental tests and analytical models, we evaluate the mechanical properties of the biomimetic material. Then, through the design of experiments and the response surface methodology, we parametrically optimize the geometry of the diatom-inspired material, maximizing its energy absorption capabilities. As a result, we obtain an architected material that not only optimizes mechanical energy absorption, but also exploits the frustule topology to improve helmet liners with enhanced breathability, abrasion resistance, lightweight, and overall multifunctionality.

## Methods

### Biomimetic design

The microstructure of *Coscinodiscus sp.* diatoms is characterized by three main layers: cribrum, areolae, and foramen (see Figure 1 (c)). Each of these layers contributes differently to the multifunctionality of the frustule [56]–[58], [62], [72], [73]. The cribrum (*i.e.*, the outer layer) prevents harmful particles or pathogens from entering the organism while allowing only essential molecules to pass through. Being thin and highly porous, it does not significantly impact the mechanical properties of the frustule but helps maintain its lightweight structure. The areolae layer (*i.e.*, the intermediate layer) is composed of hexagonal cells with a honeycomb-like arrangement. This architecture allows diatoms to maximize nutrient storage while minimizing the volume of material needed to construct their exoskeleton [74]. Additionally, it is crucial for the integrity of the frustule, as it is the most structurally efficient layer in defending diatoms against predatory attacks, which primarily result in compressive loads. The foramen (*i.e.*, the innermost layer) features reinforced holes that facilitate mass exchange between the cell and the exoskeleton chambers, and contribute to the toughness of the structure [62].

The shape of the frustule plays a crucial role in determining its mechanical and fluid-dynamic characteristics, transcending the influence of its material composition. With the specific goal of improving the multifunctionality of helmets, we therefore created a biomimetic material model that retains the intricate architecture of *Coscinodiscus sp.* diatoms but employs thermoplastic polyurethane (TPU) as the base material. Compared with the natural biosilica, TPU offers advantages such as energy absorption over multiple loading cycles and durability. In addition, because it can be processed with additive manufacturing, it facilitates the customization of helmets to meet the specific needs of users and applications. In developing this concept, we simplified the original structure by creating a periodic geometric model based on a representative volume element (RVE). Then, we adjusted the dimensions of its geometric features to the millimeter scale, making them suitable for 3D printing with Fused Filament Fabrication technology, while preserving the proportions of the biological structure (see Figure 2 (a)). Supporting Information (SI) SI Tables 1 shows the detailed dimensions of the geometric features of the frustule depicted in Figure 2 (b), for both natural [36], [62], [75]–[77] and biomimetic materials.

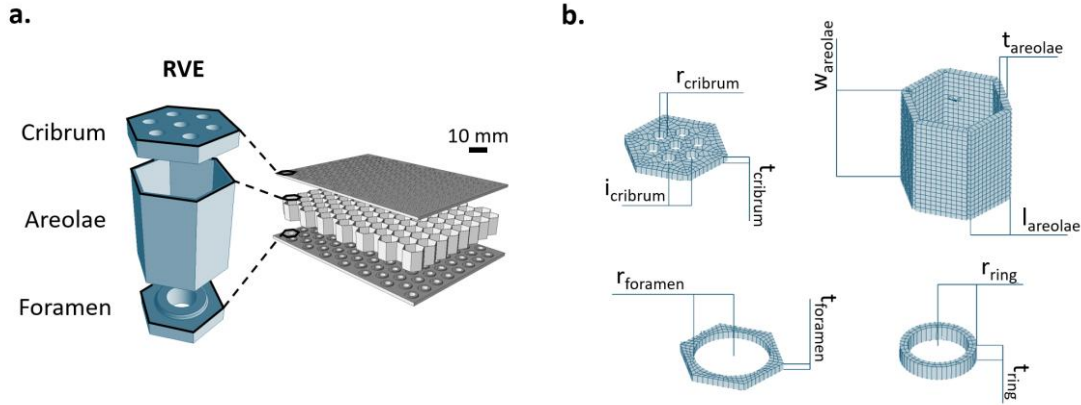

*Figure 2 – (a.) Diatom-inspired material model. For its design, we exploited a modular approach based on the spatial repetition of a Representative Volume Element (RVE) that mimics the same architecture as the Coscinodiscus sp. diatoms. (b.) Finite element model of the biomimetic RVE, broken down into its main geometric features (shell elements section properties are highlighted to emphasize the actual geometry). Significant geometric parameters were defined for each substructure. Parametric optimization was used to determine the values that maximize the energy absorption capabilities of the diatom-inspired multifunctional material.*

**Numerical simulations and parametric optimization**

We built an FE model of the RVE using ANSYS Mechanical APDL to evaluate the energy absorption properties of the biomimetic material under quasi-static out-of-plane compression. Taking advantage of the geometric, load, and material symmetry of the RVE, we discretized one-twelfth of its domain with SHELL281 structural elements (8-node quadratic shell elements) and applied symmetry boundary conditions to minimize the computational load. We then conducted a mesh convergence analysis to minimize numerical computational errors. The compression of the RVE was reproduced using two rigid contact surfaces to model the plates used in the experimental tests. The upper surface, free to move only along the axial frustule cell axis, was assigned a velocity of 2 mm/min, while the lower surface was fixed in place. Contact interactions were defined using pairs of contact-target elements and a friction coefficient of 1.2 used to simulate sliding between the rigid surfaces and the FE mesh.

We did not apply periodic boundary conditions to the RVE. This means that the proposed simulation approach, based on a single-cell analysis, cannot accurately replicate the mechanical behavior of the periodic material. However, this approach allows us to generalize our study in two ways: (i) optimizing the performance of a single unit cell is likely to lead to overall improvements in the corresponding architected material; and (ii) the results obtained are not only useful for designing helmet liners but can also be applied to other energy-absorbing devices [42].

We replicated the behavior of TPU using the hyperelastic Blatz-Ko model [78], which we calibrated based on uniaxial tensile and compression tests performed on 3D-printed samples. We did the calibration with the commercial software PolymerFEM [79]. After that, we imported the parameters of the hyperelastic model into ANSYS Mechanical APDL. Our simulations were conducted using an implicit approach with the Newton-Raphson method to calculate the nonlinear solution.

For reliable numerical results, we validated the FE model both analytically and experimentally by considering only the areolae. This approach was chosen because: (i) the out-of-plane mechanics of honeycombs can be applied, (ii) it is the layer that contributes most to the energy absorption properties of the frustule under compression, and (iii) it is a widespread structural model in energy absorption applications, thus it represents an excellent benchmark for comparison [80]. After validating the FE model, we performed a parametric optimization of the RVE geometry to maximize its energy absorption capacity. Specifically, we parameterized its geometry (see Figure 2 (b)) using an input file written in Ansys Parametric Design Language (APDL). Then, using the ANSYS Workbench environment, we optimized the model to maximize the energy per unit volume absorbed by the material:

$$U = \int_0^\varepsilon \sigma(\varepsilon)d\varepsilon \tag{1}$$

where $\sigma$ and $\varepsilon$ are the stress and strain experienced by the RVE under axial compression. Lower and upper bounds for the RVE's geometric dimensions were set based on typical measurements of natural diatom frustules, scaled to the RVE's size. Design of experiments techniques were employed to minimize the number of analyses needed to find the global optimum. Input parameter sets were selected using a Central Composite Design (CCD). This method allowed us to quantify the sensitivity of the structure's mechanical response to variations in its geometric parameters. A response surface was then interpolated over the results to identify the relative optimum.

**Analytical modeling**

The out-of-plane compression behavior of honeycombs follows the general response curve of structural foams, as shown in Figure 1 (b) [16], [81]. The energy absorbed by the material can be calculated by approximating the linear deformation and plateau phases, using the Young's modulus and the average plateau stress.

The Young's modulus of the honeycomb (i.e, the areolae layer), $E_{areolae}$, is closely related to its relative density, $\rho_{rel}$, and is given by:

$$E_{areolae} = \rho_{rel}E_{bulk} \cong \frac{2}{\sqrt{3}} \frac{t_{areolae}}{l_{areolae}} E_{bulk} \tag{2}$$

where $t_{areolae}$ and $l_{areolae}$ represent the wall thickness and the edge length (see Figure 2 (b)), respectively, and $E_{bulk}$ is the Young's modulus of the bulk material ($E_{bulk} = 150\ MPa$). The average plateau stress, on the other hand, is linked to the nature of the material that makes up the honeycomb. When the base material is hyperelastic, the plateau region begins when the hexagonal cells buckle elastically. The peak stress can then be calculated as the critical stress using the elastic instability theory of plates in compression:

$$\sigma_{cr} = K \frac{E_{bulk}}{(1 - \nu_{bulk}^2)} \rho_{rel} \left(\frac{t_{areolae}}{l_{areolae}}\right)^2 \tag{3}$$

where $K$ is a constraint factor, which depends on the boundary conditions applied to the plates constituting the honeycomb cell walls ($K = 4$) [82], and $\nu_{bulk}$ is the Poisson's coefficient of the parent material.

**3D printing and testing**

We used the Original Prusa i3 MK2s 3D printer, based on Fused Filament Fabrication (FFF) technology, to produce the specimens needed for TPU mechanical characterization and the frustule cell samples. To ensure the repeatability of results, we considered a minimum of three samples per test and geometry. The process parameters are provided in the SI. For all the samples, we used the commercial filament FILOALFA® ALFA TPU hard.

To calibrate the hyperelastic model of TPU, we conducted both tensile and compression tests according to ASTM D412 [83] and ASTM D575 [84] standards, respectively. For the honeycomb and frustule unit cells, we performed quasi-static compression tests in displacement control, using a crosshead speed of 2 mm/min, with a ZwickRoell ProLine universal testing machine equipped with a 10 kN load cell.

# Results and discussion

**Numerical model validation**

Figure 3 (a) compares results obtained for the out-of-plane compression of the bioinspired areolae unit cell (HC), both in terms of stress-strain curves and deformation modes of the structure. There is a strong correlation in terms of compression modulus, peak stress, and absorbed energy. In the elastic region, the numerical and experimental curves closely match, showing a difference of 11.3% in terms of Young's modulus. The critical stress predicted by FE analysis aligns closely with both experimental and analytical results, with errors of 2.1% and 0.8%, respectively. It is important to note that the calculation of elastic moduli is highly sensitive to the selected strain range, so the related errors are less significant compared to those for other mechanical properties. Furthermore, beyond the critical load, the FE model becomes stiffer and deviates from the experimental results. This discrepancy is mainly due to anisotropies and imperfections introduced by the additive manufacturing process in the test specimens.

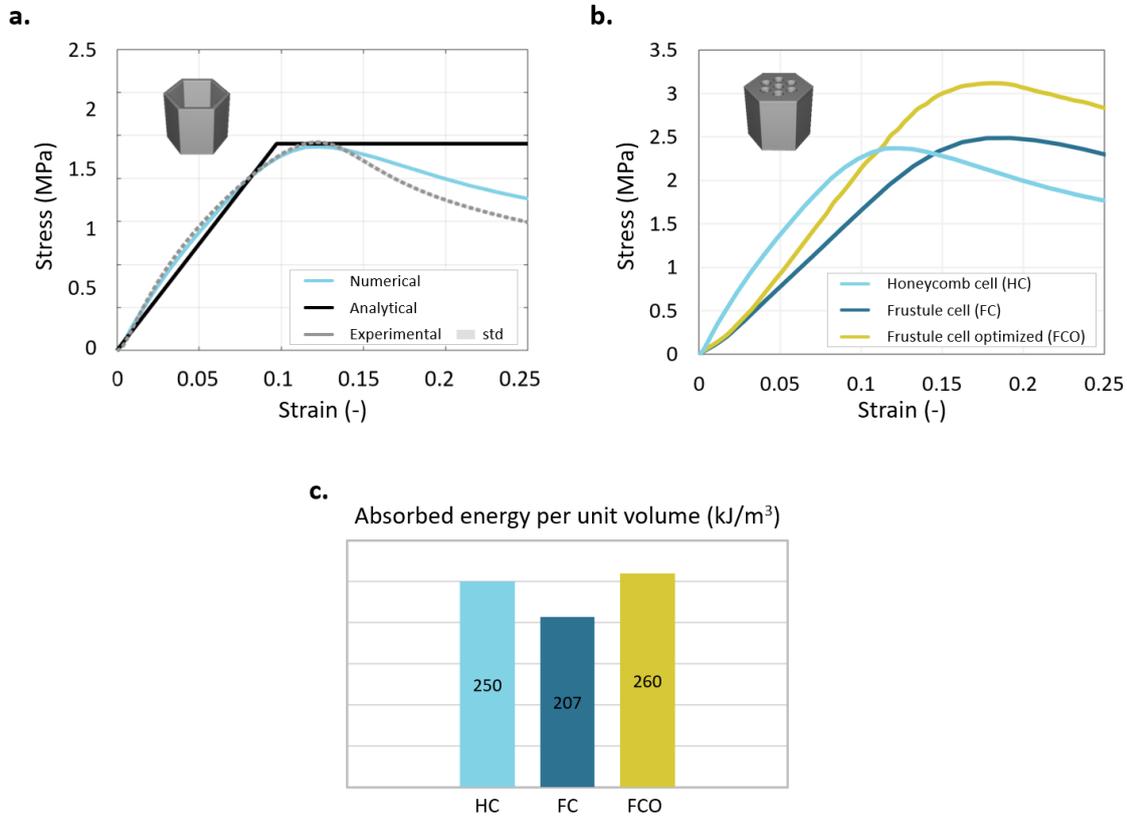

*Figure 3 – (a.) Comparison of response curves obtained analytically, numerically, and experimentally for the honeycomb cell subjected to out-of-plane compression (the small standard deviation of the experimental results indicates high repeatability). (b.) Stress-strain curves obtained by finite element analysis for the honeycomb unit cell (HC), bioinspired frustule unit cell (FC) and optimized bioinspired frustule unit cell (FCO). (c.) Energy absorption capacity of the three analyzed structures, obtained by finite element analysis .*

**Biomimetic frustule energy absorption properties**

Figure 3 (b) and Figure 3 (c) illustrate the axial compression behavior of the original (FC) and optimized (FCO) biomimetic frustule RVE compared to the honeycomb cell (HC) that models only the intermediate layer, in terms of stress-strain curves and energy absorbed per unit volume, respectively. Adding porous layers to the areolae structure of the frustule, without changing its geometric dimensions, degrades structural performance. Despite the peak stresses of both architectures being comparable, the FC cell is less rigid and absorbs less energy per unit volume through elastic deformation. In nature, the hierarchical multi-layer composition of the diatom frustule enhances the mechanical properties and provides multifunctionality. However, in the FC model we designed, adding the cribrum and foramen expands the structure's functions at the expense of energy absorption capacity. This highlights how our tools and engineering approach can fail to achieve nature's design efficiency, when mimicking simplified geometrical features. To bridge this gap, we conducted parametric optimization of the FC cell. FCO cell dimensions are given in the SI. Its response curve shows an initial stiffness that is between that of the HC and FC, but it reaches higher stresses than both structures. This indicates it can absorb more energy before buckling failure, as confirmed by the data. The optimization allows the FCO to outperform both HC and FC in terms of energy absorption. The geometric parameter ranges for the RVE were derived from those typically observed in nature, so the function of the porous layers was not significantly altered. This does not mean we have surpassed nature, as we have only replicated part of the frustule's natural functions to fit our purpose. However, it demonstrates that biomimicry is an effective approach to technological innovation.

**Effect of geometric features on frustule performance**

To gain a thorough understanding of the biomimetic frustule's structure-property relationship, we also conducted a local sensitivity analysis of the numerical response to variations in the geometric parameters of its RVE. The results are shown in Figure 4. Local sensitivity measures how a small variation in an input parameter of the optimization process affects the model's response (*i.e.*, the absorbed energy) around the nominal values in the parameter space. Positive bars indicate that an increase in the corresponding parameter leads to an increase in the objective function, while negative bars indicate an inverse proportionality. Bars with sensitivity larger than 10% (absolute value) are marked with diagonal lines to highlight the parameters that significantly influence the energy absorbed by the biomimetic frustule. . The energy absorbed per unit volume by the FC cell is mainly influenced by the geometric characteristics of the areolae, particularly its height, side length, and wall thickness. Consistent with previous observations, the honeycomb architecture proves to be highly efficient in absorbing mechanical energy and is crucial for the compressive response of the diatom-inspired material. However, it is important to note that the dimensions of the foramen and cribrum also significantly impact the overall performance of the frustule. This highlights the frustule's topology as a promising multifunctional biomimetic model for energy absorption applications. Emphasizing the combined influence of the cribrum, areolae, and foramen geometries provides a comprehensive understanding of how to optimize this structure for maximum energy absorption efficiency.

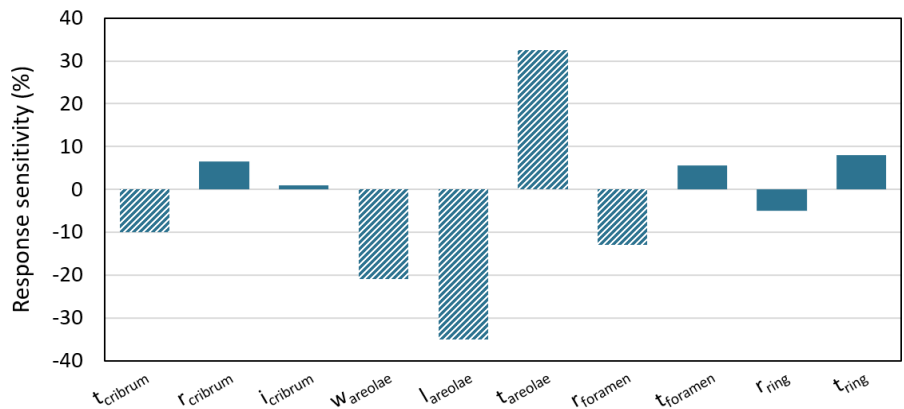

*Figure 4 - Local sensitivity analysis of the compression response of the diatom-inspired RVE to variations in its geometric parameters. Positive bars indicate a direct relationship, and negative bars indicate an inverse relationship with the absorbed energy. Parameters with sensitivity larger than 10% are marked with diagonal lines.*

# Conclusions and future perspectives

In this study, we explored the potential of biomimetic material design inspired by the frustule of *Coscinodiscus sp.* diatoms to develop advanced liners for innovative helmets. We simultaneously leverage the energy-absorbing features provided by the diatom-inspired structure, a polymeric hyperelastic material, and the parametric optimization to show how a synergistic use of biomimicry, materials design, and engineering design can lead to a novel concept, which can be fine-tuned to provide multiple functions. Through experimental tests and numerical simulations, we demonstrated that the unique topology of these microalgae, combined with engineering materials such as TPU, enable an expansion of the functionalities of conventional energy-absorption devices, without affecting safety. Specifically, we studied the mechanical response to quasi-static axial compression of a representative volume element (RVE) that mimics the architecture of diatom frustules. The results demonstrate the potential of this structure for energy absorption applications and serve as a starting point for developing innovative helmet concepts. Figure 5 illustrates a roadmap to the prototyping of a D-HAT, Diatom-inspired Helmet Against Trauma. The main phases include: (i) user-based design through reverse engineering to ensure optimal fit and comfort [85]; (ii) hierarchical structure optimization to enhance mechanical and multifunctional properties [86], (iii) homogenization of the biomimetic periodic material properties [71], [87] and safety approval through explicit dynamic FE simulations and impact tests, and (iv) the production of the final product via 3D printing. Future developments will include integrating smart functionalities into the diatom-inspire structure. For instance, incorporating conductive nanofillers into the TPU matrix could enable real-time structural health monitoring capabilities, leading to more informed maintenance and replacement decisions [88].

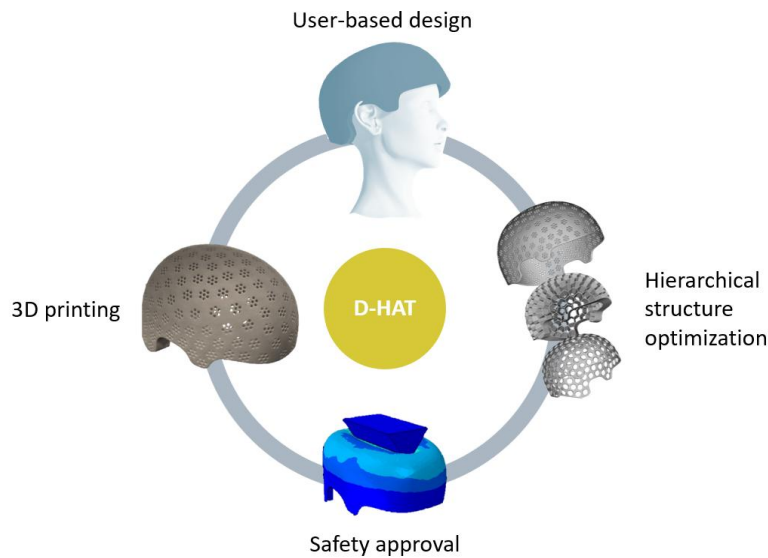

*Figure 5 – A roadmap to the prototyping of a Diatom-inspired Helmet Against Trauma (D-HAT). Main steps include helmet user-based design through reverse engineering, lattice-based hierarchical structure optimization of the liner, material homogenization for simulations and safety approval, and 3D printing of the final product.*

## Acknowledgements

The authors acknowledge Dr. Anna Pippo for developing the aesthetic design and renderings of the diatom-inspired helmet. The authors acknowledge support from the University of Genoa, under the Curiosity Driven Starting Grant initiative by NextGEneration EU.

# Supporting Information

## RVE dimensions

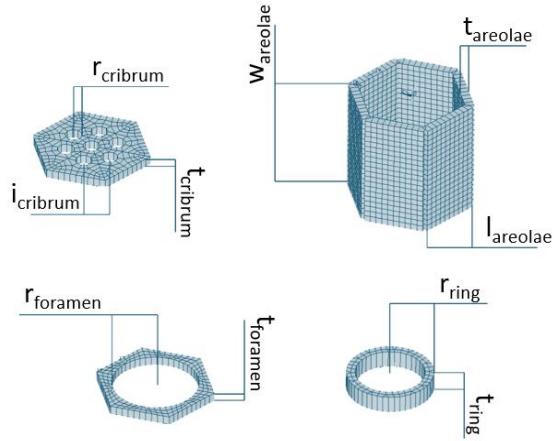

*SI Figure 1 - Geometric features of the diatom frustule RVE.*

*SI Table 0.1 – Geometric feature dimensions of the biological [1]–[5] and bioinspired Representative Volume Element of the Coscinodiscus sp. diatom frustule. The nominal values of the bioinspired model were defined by choosing a central value within the ranges of values measured in the literature and applying a magnification factor of 7000x.*

| **Cribrum layer** | | | |
|---|---|---|---|
| | Thickness ($t_{cribrum}$) | Hole radius ($r_{cribrum}$) | Hole spacing ($i_{cribrum}$) |
| Natural structure | 250 ÷ 270 nm | 200 ÷ 300 nm | ~ |
| Biomimetic structure | 2 mm | 2 mm | 6 mm |
| **Areolae layer** | | | |
| | Wall height ($w_{areolae}$) | Wall thickness ($t_{areolae}$) | Wall side ($l_{areolae}$) |
| Natural structure | 3442 ÷ 4938 nm | 280 ÷ 290 nm | 1000 ÷ 3000 nm |
| Biomimetic structure | 28 mm | 2 mm | 14 mm |
| **Foramen layer** | | | |
| | Thickness ($t_{foramen}$) | Hole radius ($r_{foramen}$) | Ring radius ($r_{ring}$) | Ring thickness ($t_{ring}$) |
| Natural structure | 360 ÷ 420 nm | 1121 ÷ 1180 nm | 1370 ÷ 1480 nm | 280 ÷ 290 nm |
| Biomimetic structure | 2 mm | 8 mm | 10 mm | 2 mm |

## 3D Printing

*SI Table 2 - 3D printing process parameters used for TPU specimens.*

| **Process parameter** | **Unit** | **Value** |
|---|---|---|
| Extruder temperature | °C | 230 |
| Max printing speed | mm/s | 40 |

| Printer bed temperature | °C | 50 |
|---|---|---|
| Drying temperature | °C | 70 |
| Drying time | hours | 5 |
| Infill percentage | % | 100 |
| Other parameters | - | Prusa slicer default values |

# Parametric optimization results

*SI Table 3 – Optimized dimensions of the biomimetic RVE, identified as Frustule Cell Optimized (FCO).*

| **Cribrum layer** | | | |
|---|---|---|---|
| | Thickness ($t_{cribrum}$) | Hole radius ($r_{cribrum}$) | Hole spacing ($i_{cribrum}$) |
| FCO | 1.5 mm | 2.33 mm | 7 mm |
| **Areolae layer** | | | |
| | Wall height ($w_{areolae}$) | Wall thickness ($t_{areolae}$) | Wall side ($l_{areolae}$) |
| FCO | 25 mm | 2.3 mm | 14 mm |
| **Foramen layer** | | | |
| | Thickness ($t_{foramen}$) | Hole radius ($r_{foramen}$) | Ring radius ($r_{ring}$) | Ring thickness ($t_{ring}$) |
| FCO | 2.44 mm | 8.65 mm | 10 mm | 2.3 mm |

# Supporting Information references